\pdfoutput=1

\documentclass[11pt]{article}

\usepackage{acl}
\usepackage{times}
\usepackage{latexsym}

\usepackage[T1]{fontenc}

\usepackage[utf8]{inputenc}

\usepackage{microtype}
\usepackage[ruled,vlined]{algorithm2e}
\usepackage{inconsolata}
\usepackage{multirow}
\usepackage{graphicx}

\title{Debate-Feedback: A Multi-Agent Framework for Efficient Legal Judgment Prediction }

\author{Xi Chen \\
  \texttt{xich0108@bu.edu} \\\And
  Mao Mao \\
  \texttt{maomao@bu.edu} \\\And
  Shuo Li \\
  \texttt{lis23@m.fudan.edu.cn} \\\And
  Haotian Shangguan \\
  \texttt{haosg19@bu.edu} \\
  }

\usepackage{amsmath}
\usepackage{booktabs}
\usepackage{hyperref}
\usepackage{tabularx} 
\usepackage{adjustbox}
\usepackage{longtable}
\usepackage{colortbl}
\begin{document}
\maketitle
\begin{abstract}
The use of AI in legal analysis and prediction (LegalAI) has gained widespread attention, with past research focusing on retrieval-based methods and fine-tuning large models. However, these approaches often require large datasets and underutilize the capabilities of modern large language models (LLMs). In this paper, inspired by the debate phase of real courtroom trials, we propose a novel legal judgment prediction model based on the Debate-Feedback architecture, which integrates LLM multi-agent debate and reliability evaluation models. Unlike traditional methods, our model achieves significant improvements in efficiency by minimizing the need for large historical datasets, thus offering a lightweight yet robust solution. Comparative experiments show that it outperforms several general-purpose and domain-specific legal models, offering a dynamic reasoning process and a promising direction for future LegalAI research. 
\end{abstract}
\section{Introduction}
\begin{figure*}[t]
    \centering
    \includegraphics[width=12cm, height=4.5cm]{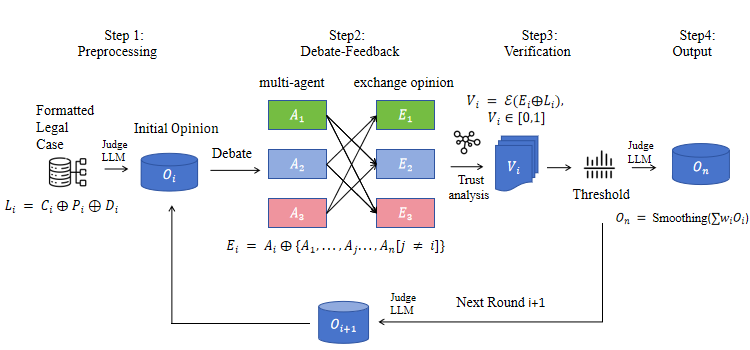}
    \caption{A brief introduction of Debate-Feedback Structure}
    \label{fig:Debate-Feedback Model}
\end{figure*}
LegalAI leverages artificial intelligence technologies such as natural language processing, machine learning, and deep learning to address various legal tasks \citep{aletras2016predicting,article,zhong2020how}, including legal document analysis and consultation. A key area of LegalAI is Legal Judgment Prediction (LJP) \citep{zhong-etal-2018-legal,ma2021precedent,10255647}, which focuses on predicting court judgments. LJP tasks typically use historical legal case data, including background information, case descriptions, statements from both parties, precedents, and court verdicts. Predictions range from binary outcomes (e.g., plaintiff vs. defendant wins) to multi-class tasks (e.g., sentence prediction). NLP technologies, combined with advanced models like LegalBERT \cite{chalkidis2019legal} and Lawformer \cite{xiao2021lawformerpretrainedlanguagemodel}, have achieved strong results by learning from large datasets.

The debate model is a system that integrates large language modeling (LLM) with argumentative reasoning techniques to simulate the process of debate or contention \citep{irving2018ai,nie2020adversarialnlinewbenchmark}, ultimately arriving at a decision or conclusion on a specific issue through the debate process. In a typical debate task, multiple LLM agents assume different roles and are deliberately guided to provide answers from various perspectives or positions. These generated arguments are then synthesized to assist the LLM in reaching a final conclusion \cite{zeng2022socraticmodelscomposingzeroshot}.

In this paper, we propose a Debate-Feedback model to explore an efficient and convenient method for predicting legal judgement. Fig[\ref{fig:Debate-Feedback Model}] shows the general framework of the model in the task of predicting decision results. Specifically, Debate-Feedback can be divided into four steps. First, the collected historical legal cases $L_i$ will be formatted into Case Background $C_i$, Plaintiff Claim $P_i$ and Defendant Statement $D_i$. These information will be provided to the judge LLM for initial prediction. In the second step of the debate, multiple LLM agents will be guided to answer the prediction questions from different perspectives, and then exchange opinions and debate to generate their own comments $E_i$. In the verification phase, a pre-trained assistant model $\mathcal{E}$ will conduct a reliability analysis on each LLM's comments combined with case information. The results of the analysis will be provided to the judge LLM for reference together with each agent's comments. The judge LLM will give the prediction $O_i$ for this round based on the above information $\mathcal{E} = E_i \oplus L_i$. More details are illustrated in the Methodology section. In summary, we introduce a Debate-Feedback model that enhances legal judgment prediction by incorporating a multi-agent debate process and reliability evaluation, providing a more efficient and accurate solution with reduced reliance on large datasets.

\section{Related Work}
Legal documents are characterized by lengthy texts and complex logic, which has led prior research to focus on two key approaches to address these challenges: training legal LLM and using retrieval augmentation.
\subsection{Legal LLM}
In-context Learning(ICL) is a learning paradigm widely applied in large language models (LLMs) by using a set of context examples to guide predictions during reasoning \citep{dong2024surveyincontextlearning,liu2021pretrainpromptpredictsystematic,gutierrezpachas2022distributionmarkovchainreversetime,min2022metaicllearninglearncontext}. However, due to the often extensive length of legal texts, naive ICL methods are constrained by LLM input length limits. As a result, LegalAI solutions typically combine ICL with fine-tuning or pre-training of models to overcome these limitations. For instance, LegalBERT \cite{chalkidis2019legal} fine-tunes BERT on legal datasets, achieving strong results in legal text classification and provision retrieval. Similarly, Lawformer \cite{xiao2021lawformerpretrainedlanguagemodel} handles lengthy Chinese legal documents, while CaseLaw-BERT \cite{paul2023pretrainedlanguagemodelslegal}, fine-tuned on case law datasets, enhances legal case retrieval and judgment prediction. Despite their success, these approaches rely heavily on large, domain-specific datasets, which can limit their applicability across different legal systems and languages.

\subsection{Retrieval Augmentation}
Retrieving relevant legal precedents—court judgments or legal decisions from previous cases—is a mainstream approach to assist LLMs in making predictions, especially in overcoming the challenge of lengthy texts. By providing recommended samples, this method guides the LLM's reasoning process more effectively \citep{zhong2020how,10.1145/3462757.3466066}. Ma et al. introduced a framework that deeply integrates legal precedents into judgment prediction \cite{wu-etal-2023-precedent}, combining the reasoning capabilities of LLMs with domain-specific models to enable more accurate and context-aware predictions. Similarly, Caseformer \cite{su2024caseformerpretraininglegalcase} employs a pre-training strategy that emphasizes distinctions between cases, enhancing case retrieval performance. Although retrieval augmentation improves the handling of long texts, it still relies on the availability of large datasets, and its reliance on specific legal systems and languages can limit broader applicability across different jurisdictions.

\section{Methodology}
In this section, we first systematically introduce our Feedback-Debate model, followed by an analysis of the limitations of the general debate architecture in specific legal scenarios, along with proposed solutions to address these shortcomings. 

\textbf{Overview}
Algorithm[\ref{tab: Algorithm}] presents the pseudo code for the debate-feedback framework in binary classification. The input is a preprocessed legal event text, labeled as $S$, and the main language model (LM) plays the role of the judge, predicting the probability of a legal judgment, $LM: S \rightarrow [0,1]$. Two agents, $t_{ne}$ and $t_{po}$, debate from opposing perspectives, providing inputs to refine the judgment. Each debate round involves these agents exchanging and debating their positions, with $n$ defining the number of iterations.

The assistant model $\mathcal{E}$ evaluates the reliability of the agents' arguments and outputs a probability. If the reliability exceeds a threshold, the main LM adjusts its prediction by weighting the latest information, otherwise it defaults to the initial prediction. The final decision is smoothed over all rounds to produce a stable outcome.
(Note that notation $\oplus$ does not mean xor, but rather combination in a non-additive sense.)

\begin{algorithm}[t]
\SetAlgoLined
\KwIn{ 
  LM, $\mathcal{E}$ : $\mathcal{S} \to [0, 1]$; $n$, $T$ $\in$ $N$; $x \in S$ ;
  $t_{ne}$, $t_{po}$ : $\mathcal{S} \to \mathcal{S}$;
}
\KwOut{Final decision $y \in (0,1)$;}

$O_0 \leftarrow LM(x)$;

\For{$i \leftarrow 1$ \KwTo $n$}{
    \tcp{Debate Step}
    $a: a_{ne}, a_{po} \leftarrow t_{ne}(x), t_{po}(x)$\;
    $e: e_{ne}, e_{po} \leftarrow t_{ne}(x \oplus a_{po}), t_{po}(x \oplus a_{ne})$\;
    \tcp{Verification Step}
    $v: v_{ne}, v_{po}  \leftarrow \mathcal{E}(e_{ne}), 
    \mathcal{E}(e_{po})$\;
    \quad\quad\quad $sum = LM(a,e,v)$\;
    \If {Threshold(v)}{
        $O_i = (1-T) * O_{i-1} + T*LM(x,sum)$\;
        }
    \Else{
        $O_i = LM(x)$\;
        }
}
$ y \leftarrow O_n $;

\caption{Debate-Feedback}
\label{tab: Algorithm}
\end{algorithm}

\textbf{Reliability Analysis}
Through experiments, we observe that a simple debate model can sometimes lead to worse prediction results. This occurs because legal predictions differ from mathematical problems, as they often involve subjective tendencies. A straightforward example is when we guide multiple LLMs to debate from the perspectives of the plaintiff and defendant, it is challenging for them to reach a consensus. To address this issue, one of our solutions is to train an assistant model that learns from a large corpus of legal event annotations and assists in evaluating the reliability of different debate arguments, as shown in Table[\ref{tab: assistant model}]. Specifically, the training set for the assistant model is generated from multiple runs of the unassisted Debate-Feedback model, which we refer to as Debate-Feedback (single) in the subsequent experimental section.
\begin{table}[]
\begin{adjustbox}{max width=\linewidth}
\begin{tabular}{|l|l|}
\hline
\multicolumn{2}{|l|}{\textbf{TrainingSet of Assistant model}}                   \\ \hline
\textbf{Training\_X} & \{Case\_background + Debater's opinion\} \\ \hline
\textbf{Training\_Y} & \{Ground\_truth \textbf{XOR} Debater's position\}    \\ \hline

\end{tabular}
\end{adjustbox}
\caption{Dataset of assistant model.}
\label{tab: assistant model}
\end{table}

\textbf{Smoothing Operation}
To mitigate the impact of a "failed" debate where the main LLM generates incorrect answers, we apply a smoothing operation. This involves saving the results of each prediction and assigning them a certain weight. Specifically, let $LM(x)$ represent the predicted result of the $i$-th debate and $T$ be the weighting factor. The updated result is calculated as:
\begin{equation}
O_i \leftarrow (1-T) * O_{i-1} + T * LM(x)
\label{eq:smoothing}
\end{equation}
where $T \in [0, 1]$ represents the weight assigned to the latest prediction.


\section{Experiment}
\subsection{Dateset and Baseline}

Along with many influential LegalAI works, we also use CaseLaw as the main dataset. The \textbf{CaseLaw} dataset is a legal case dataset specifically used for natural language processing (NLP) and machine learning tasks in the legal field, especially in the fields of legal case retrieval and legal judgment prediction. This dataset contains a large number of court case texts that have been judged, usually including descriptions of legal facts, legal reasoning, and judgment results. In order to test the model's cross-language and cross-legal capabilities, we also used the Chinese dataset \textbf{CAIL18}~\citep{xiao2018cail2018largescalelegaldataset,zhong2018overviewcail2018legaljudgment}.

We compare Debate-Feedback with both general large language models and legal domain models. \textbf{GPT4o} and \textbf{GPT3.5-turbo} are representative general large language models at present \cite{openai2024gpt4technicalreport}, and they have been proven to have strong text analysis and logical reasoning capabilities. \textbf{LegalBert} \cite{chalkidis2019legal} and \textbf{Lawformer} \cite{xiao2021lawformerpretrainedlanguagemodel} are well-known legal domain model, they're able to capture the association between legal terms and cases well. In addition, \textbf{CNN} \cite{726791} is also used as a classifier for feature extraction in the baseline evaluation, with \textbf{BERT} \cite{devlin2019bertpretrainingdeepbidirectional} serving as the text embedding layer.

Considering that the debate-feedback framework can essentially be seen as a large language model reasoning framework, we also compare it with classic reasoning methods, including \textbf{Few-shot Learning}, \textbf{Chain of Thought(CoT)} \cite{wei2023chainofthoughtpromptingelicitsreasoning} and \textbf{Reflexion} \cite{shinn2023reflexionlanguageagentsverbal}. We use gpt-4o mini as the baseline model in this part and verified them on a smaller subset on a smaller subset of the datasets (12,000 samples from CaseLaw and 3,000 samples from CAIL18). 

\subsection{Regular LJP tasks}

\textbf{Trial Prediction}
The input for trial prediction includes a legal text, along with the opinions of the plaintiff and defendant. The predicted labels are Plaintiff wins, Defendant wins, Settlement, and Dismissed. Since Settlement and Dismissed are explicitly stated in the legal text, this can be reduced to a binary classification task with two labels: Plaintiff wins and Defendant wins. The CaseLaw dataset was used for this task, and Table[\ref{tab: sample of DF model}] provides a sample.
\newline
\textbf{Article Prediction}
Article prediction is a multi-label classification task. The model receives a description of legal facts and the prediction content contains multiple labels of different relevant law articles. CAIL18 dataset is used in this task.

\subsection{Evaluation Metrics}

In this study, we evaluate the model performance using two key metrics: accuracy and F1-score.

\textbf{Accuracy(Acc)} is the proportion of correct predictions among all predictions. It is computed as:

\begin{equation}
\text{Accuracy} = \frac{\sum_{i=1}^{N} (y_i = y_{\text{true},i})}{N}
\end{equation}

where $N$ is the total number of predictions, $y_i$ is the predicted label, $y_{\text{true},i}$ is the actual label, and $(\cdot)$ is the indicator function that equals 1 when the condition is true and 0 otherwise.

\textbf{F1-score(F1)} is useful for imbalanced datasets as it balances precision and recall. In multi-class classification, F1-score is computed for each class and then averaged (macro F1-score). For a single class, F1-score is given by:

\begin{equation}
F1 = 2 \times \frac{\text{Precision} \times \text{Recall}}{\text{Precision} + \text{Recall}}
\end{equation}

Where precision and recall are defined as:

\begin{equation}
\text{Precision} = \frac{\sum_{i=1}^{N} {1}(y_i = c \land y_{\text{true},i} = c)}{\sum_{i=1}^{N} {1}(y_i = c)}
\end{equation}

\begin{equation}
\text{Recall} = \frac{\sum_{i=1}^{N} {1}(y_i = c \land y_{\text{true},i} = c)}{\sum_{i=1}^{N} {1}(y_{\text{true},i} = c)}
\end{equation}

For multi-class classification, the macro F1-score is calculated as the average F1-scores for all classes:

\begin{equation}
F1_{\text{macro}} = \frac{1}{C} \sum_{c=1}^{C} F1_c
\end{equation}

where $C$ is the number of classes.

\subsection{Experimental Results}
\begin{table}[]
\centering
\scalebox{0.85}{
\begin{tabular}{|l|l|l|l|l|}
\hline
\multirow{2}{*}{Model} & \multicolumn{2}{c|}{CaseLaw} & \multicolumn{2}{c|}{CAIL18} \\ \cline{2-5} 
                       & Acc           & F1           & Acc           & F1          \\ \hline
CNN(with BERT)                    &    0.58           &     0.54         &    0.39           &    0.11         \\ 
Legal-BERT             & 0.63          &    0.61          &       0.22        &      0.03       \\ 
Lawformer             &   0.53       &      0.31       &         0.38      &     0.12       \\ 
GPT-3.5-turbo          & 0.49          &    0.27          & 0.26          & 0.04        \\ 
GPT-4o               & 0.64          &     0.64         & 0.31          & 0.05        \\ \hline
Debate-Feedback(single)        & 0.66          &    0.65          & 0.42          & \textbf{0.16}        \\ 
Debate-Feedback(assistant)        & \textbf{0.67}          &    \textbf{0.66}          & \textbf{0.45}          & \textbf{0.16}        \\ \hline
\end{tabular}
}
\caption{Comparison of models on CaseLaw and CAIL18 datasets. All judge's and debaters' LMs in experiments are based on the GPT-4o model and T = 0.5.}
\label{tab:result}
\end{table}

The experimental results demonstrate the effectiveness of the Debate-Feedback model, with the inclusion of an assistant model in the feedback loop enhancing prediction reliability and providing more robust results compared to the single Debate-Feedback model. These results validate the strength of our approach in improving the accuracy and consistency of legal judgment predictions. Our experimental results are shown in Table[\ref{tab:result}], Figure[\ref{fig:numberofdebaters}] and Figure[\ref{fig:numberofrounds}].

\textbf{CaseLaw Dataset Performance}
For the CaseLaw dataset, the Debate-Feedback model outperformed GPT-4o, GPT-3.5-turbo, Legal-BERT, CNN and Lawformer. The model with the assistant achieved an accuracy of 0.67 and an F1-score of 0.66, while the single Debate-Feedback model obtained slightly lower performance with an accuracy of 0.66 and an F1-score of 0.65. These results show that our method improves the performance of pre-train legal domain models, which only achieved an accuracy of 0.63 and an F1-score of 0.61. The assistant model's inclusion in the feedback loop improves the reliability of predictions, making it more robust compared to the single model.

\textbf{CAIL18 Dataset Performance}
On the Chinese legal dataset CAIL18, the Debate-Feedback model achieved a remarkable accuracy of 0.45, significantly surpassing GPT-4o (accuracy 0.31) and GPT-3.5-turbo (accuracy 0.26). The model with an assistant component further improved the F1-score to 0.16, highlighting the ability of the assistant model to refine predictions and correct any inconsistencies in the debate phase. These results also suggest that the Debate-Feedback model is more versatile in handling cross-linguistic challenges compared to other models.

\begin{table}[h]
    \centering
    \scalebox{0.75}{
    \begin{tabular}{|l|l|l|l|l|}
        \hline
        \multirow{2}{*}{Model} & \multicolumn{2}{c|}{CaseLaw} & \multicolumn{2}{c|}{CAIL18} \\
        \cline{2-5} 
        & Acc & F1 & Acc & F1 \\ \hline
        \midrule
        Few-shot & 63.8\% & 64.1\% & 29.7\% & 5.03\% \\
        CoT (4-steps) & 63.7\% & 64.0\% & 31.2\% & 6.17\% \\
        Reflexion & 64.5\% & 65.0\% & 31.8\% & 8.12\% \\ \hline
        Debate-Feedback (single) & \textbf{66.2\%} & \textbf{65.7\%} & \textbf{41.9\%} & 16.1\% \\
        Debate-Feedback (assistant) & \textbf{67.1\%} & \textbf{66.1\%} & \textbf{44.8\%} & \textbf{16.3\%} \\ \hline
        \bottomrule
    \end{tabular}
    }
    \caption{Performance comparison of different reasoning methods on CaseLaw and CAIL18 datasets.}
    \label{tab:model_comparison}
\end{table}

\textbf{Comparison with basic reasoning methods}
As shown in table[\ref{tab:model_comparison}], Debate-Feedback structure achieves significant advantages in comparison with several basic reasoning frameworks. The results show that Chain-of-Thought and Reflection perform only marginally better than Zeroshot, while our Debate-feedback framework consistently demonstrates superior performance, reinforcing the conclusions of our original experiments.

We believe there are two primary reasons why standard reasoning techniques like CoT and Reflection are less effective for this type of legal prediction problem:

\textbf{Complexity of Legal Texts}: The legal text itself is lengthy and logically complex, and simple prompts are difficult to be effective.

\textbf{Nature of Legal Prediction}: Legal prediction is always different from logical reasoning. It is not a step-by-step thinking toward the correct answer, but usually a discussion to unify or compromise the views of multiple parties. This is precisely why we designed the Debate-feedback framework, which is tailored to handle such tasks.

\section{Conclusion}
We propose a debate-feedback model based on LLMs for legal judgment prediction and demonstrated its feasibility through experiments. The inclusion of an assistant model and reliability analysis enhances prediction robustness. Future work could explore the application of debate models in other fields or further integrate them with LLMs.

\section{Limitations}
Our work currently has the following limitations:

(a) The experiments were limited to two datasets and two specific tasks, broader evaluations across additional datasets and tasks are necessary to fully validate the model’s robustness and generalizability in different legal contexts.

(b) While the smoothing technique and assistant model (reliability analysis) were included in the framework, their individual contributions to the overall performance were not deeply investigated.

(c) This work does not integrate retrieval argument techniques, which presents a promising direction for future research to enhance the model's performance.

\bibliography{main}

\begin{thebibliography}{26}
\providecommand{\natexlab}[1]{#1}

\bibitem[{Aletras et~al.(2016)Aletras, Tsarapatsanis, Preoțiuc-Pietro, and Lampos}]{aletras2016predicting}
Nikolaos Aletras, Dimitrios Tsarapatsanis, Daniel Preoțiuc-Pietro, and Vasileios Lampos. 2016.
\newblock \href {https://doi.org/10.7717/peerj-cs.93} {Predicting judicial decisions of the european court of human rights: A natural language processing perspective}.
\newblock \emph{PeerJ Computer Science}, 2:e93.

\bibitem[{Chalkidis et~al.(2019)Chalkidis, Fergadiotis, Malakasiotis, Aletras, and Androutsopoulos}]{chalkidis2019legal}
Ilias Chalkidis, Michael Fergadiotis, Prodromos Malakasiotis, Nikolaos Aletras, and Ion Androutsopoulos. 2019.
\newblock \href {https://aclanthology.org/2020.findings-emnlp.261} {Legalbert: The muppets straight out of law school}.
\newblock In \emph{Proceedings of the 2020 Conference on Empirical Methods in Natural Language Processing: Findings}, pages 2898--2904.

\bibitem[{Cui et~al.(2023)Cui, Shen, and Wen}]{10255647}
Junyun Cui, Xiaoyu Shen, and Shaochun Wen. 2023.
\newblock \href {https://doi.org/10.1109/ACCESS.2023.3317083} {A survey on legal judgment prediction: Datasets, metrics, models and challenges}.
\newblock \emph{IEEE Access}, 11:102050--102071.

\bibitem[{Devlin et~al.(2019)Devlin, Chang, Lee, and Toutanova}]{devlin2019bertpretrainingdeepbidirectional}
Jacob Devlin, Ming-Wei Chang, Kenton Lee, and Kristina Toutanova. 2019.
\newblock \href {https://arxiv.org/abs/1810.04805} {Bert: Pre-training of deep bidirectional transformers for language understanding}.
\newblock \emph{Preprint}, arXiv:1810.04805.

\bibitem[{Dong et~al.(2024)Dong, Li, Dai, Zheng, Ma, Li, Xia, Xu, Wu, Liu, Chang, Sun, Li, and Sui}]{dong2024surveyincontextlearning}
Qingxiu Dong, Lei Li, Damai Dai, Ce~Zheng, Jingyuan Ma, Rui Li, Heming Xia, Jingjing Xu, Zhiyong Wu, Tianyu Liu, Baobao Chang, Xu~Sun, Lei Li, and Zhifang Sui. 2024.
\newblock \href {https://arxiv.org/abs/2301.00234} {A survey on in-context learning}.
\newblock \emph{Preprint}, arXiv:2301.00234.

\bibitem[{Gutierrez-Pachas et~al.(2022)Gutierrez-Pachas, Costa, and Vargas}]{gutierrezpachas2022distributionmarkovchainreversetime}
Daniel~A. Gutierrez-Pachas, Eduardo~F. Costa, and Alessandro~N. Vargas. 2022.
\newblock \href {https://arxiv.org/abs/2206.05607} {Distribution of a markov chain in reverse-time with cluster observations in the extremes of a finite time window}.
\newblock \emph{Preprint}, arXiv:2206.05607.

\bibitem[{Huang et~al.(2021)Huang, Low, Teng, Zhang, Ho, Krass, and Grabmair}]{10.1145/3462757.3466066}
Zihan Huang, Charles Low, Mengqiu Teng, Hongyi Zhang, Daniel~E. Ho, Mark~S. Krass, and Matthias Grabmair. 2021.
\newblock \href {https://doi.org/10.1145/3462757.3466066} {Context-aware legal citation recommendation using deep learning}.
\newblock In \emph{Proceedings of the Eighteenth International Conference on Artificial Intelligence and Law}, ICAIL '21, page 79–88, New York, NY, USA. Association for Computing Machinery.

\bibitem[{Irving et~al.(2018)Irving, Christiano, and Amodei}]{irving2018ai}
Geoffrey Irving, Paul Christiano, and Dario Amodei. 2018.
\newblock \href {https://arxiv.org/abs/1805.00899} {Ai safety via debate}.
\newblock \emph{arXiv preprint arXiv:1805.00899}.

\bibitem[{Katz et~al.(2017)Katz, Bommarito, and Blackman}]{article}
Daniel Katz, Michael Bommarito, and Josh Blackman. 2017.
\newblock \href {https://doi.org/10.1371/journal.pone.0174698} {A general approach for predicting the behavior of the supreme court of the united states}.
\newblock \emph{PLOS ONE}, 12.

\bibitem[{Lecun et~al.(1998)Lecun, Bottou, Bengio, and Haffner}]{726791}
Y.~Lecun, L.~Bottou, Y.~Bengio, and P.~Haffner. 1998.
\newblock \href {https://doi.org/10.1109/5.726791} {Gradient-based learning applied to document recognition}.
\newblock \emph{Proceedings of the IEEE}, 86(11):2278--2324.

\bibitem[{Liu et~al.(2021)Liu, Yuan, Fu, Jiang, Hayashi, and Neubig}]{liu2021pretrainpromptpredictsystematic}
Pengfei Liu, Weizhe Yuan, Jinlan Fu, Zhengbao Jiang, Hiroaki Hayashi, and Graham Neubig. 2021.
\newblock \href {https://arxiv.org/abs/2107.13586} {Pre-train, prompt, and predict: A systematic survey of prompting methods in natural language processing}.
\newblock \emph{Preprint}, arXiv:2107.13586.

\bibitem[{Ma et~al.(2021)Ma, Liu, Wei, and Huang}]{ma2021precedent}
Jiayuan Ma, Chao Liu, Furu Wei, and Deheng Huang. 2021.
\newblock \href {https://aclanthology.org/2021.emnlp-main.368/} {Precedent-enhanced legal judgment prediction with llm and domain-model collaboration}.
\newblock In \emph{Proceedings of the 2021 Conference on Empirical Methods in Natural Language Processing}, pages 4562--4571.

\bibitem[{Min et~al.(2022)Min, Lewis, Zettlemoyer, and Hajishirzi}]{min2022metaicllearninglearncontext}
Sewon Min, Mike Lewis, Luke Zettlemoyer, and Hannaneh Hajishirzi. 2022.
\newblock \href {https://arxiv.org/abs/2110.15943} {Metaicl: Learning to learn in context}.
\newblock \emph{Preprint}, arXiv:2110.15943.

\bibitem[{Nie et~al.(2020)Nie, Williams, Dinan, Bansal, Weston, and Kiela}]{nie2020adversarialnlinewbenchmark}
Yixin Nie, Adina Williams, Emily Dinan, Mohit Bansal, Jason Weston, and Douwe Kiela. 2020.
\newblock \href {https://arxiv.org/abs/1910.14599} {Adversarial nli: A new benchmark for natural language understanding}.
\newblock \emph{Preprint}, arXiv:1910.14599.

\bibitem[{OpenAI et~al.(2024)OpenAI, Achiam, Adler, Agarwal, Ahmad, Akkaya et~al.}]{openai2024gpt4technicalreport}
OpenAI, Josh Achiam, Steven Adler, Sandhini Agarwal, Lama Ahmad, Ilge Akkaya, et~al. 2024.
\newblock \href {https://arxiv.org/abs/2303.08774} {Gpt-4 technical report}.
\newblock \emph{Preprint}, arXiv:2303.08774.

\bibitem[{Paul et~al.(2023)Paul, Mandal, Goyal, and Ghosh}]{paul2023pretrainedlanguagemodelslegal}
Shounak Paul, Arpan Mandal, Pawan Goyal, and Saptarshi Ghosh. 2023.
\newblock \href {https://arxiv.org/abs/2209.06049} {Pre-trained language models for the legal domain: A case study on indian law}.
\newblock \emph{Preprint}, arXiv:2209.06049.

\bibitem[{Shinn et~al.(2023)Shinn, Cassano, Berman, Gopinath, Narasimhan, and Yao}]{shinn2023reflexionlanguageagentsverbal}
Noah Shinn, Federico Cassano, Edward Berman, Ashwin Gopinath, Karthik Narasimhan, and Shunyu Yao. 2023.
\newblock \href {https://arxiv.org/abs/2303.11366} {Reflexion: Language agents with verbal reinforcement learning}.
\newblock \emph{Preprint}, arXiv:2303.11366.

\bibitem[{Su et~al.(2024)Su, Ai, Wu, Ma, Li, Liu, Wu, and Zhang}]{su2024caseformerpretraininglegalcase}
Weihang Su, Qingyao Ai, Yueyue Wu, Yixiao Ma, Haitao Li, Yiqun Liu, Zhijing Wu, and Min Zhang. 2024.
\newblock \href {https://arxiv.org/abs/2311.00333} {Caseformer: Pre-training for legal case retrieval based on inter-case distinctions}.
\newblock \emph{Preprint}, arXiv:2311.00333.

\bibitem[{Wei et~al.(2023)Wei, Wang, Schuurmans, Bosma, Ichter, Xia, Chi, Le, and Zhou}]{wei2023chainofthoughtpromptingelicitsreasoning}
Jason Wei, Xuezhi Wang, Dale Schuurmans, Maarten Bosma, Brian Ichter, Fei Xia, Ed~Chi, Quoc Le, and Denny Zhou. 2023.
\newblock \href {https://arxiv.org/abs/2201.11903} {Chain-of-thought prompting elicits reasoning in large language models}.
\newblock \emph{Preprint}, arXiv:2201.11903.

\bibitem[{Wu et~al.(2023)Wu, Zhou, Liu, Lu, Liu, Zhang, Sun, Wu, and Kuang}]{wu-etal-2023-precedent}
Yiquan Wu, Siying Zhou, Yifei Liu, Weiming Lu, Xiaozhong Liu, Yating Zhang, Changlong Sun, Fei Wu, and Kun Kuang. 2023.
\newblock \href {https://doi.org/10.18653/v1/2023.emnlp-main.740} {Precedent-enhanced legal judgment prediction with {LLM} and domain-model collaboration}.
\newblock In \emph{Proceedings of the 2023 Conference on Empirical Methods in Natural Language Processing}, pages 12060--12075, Singapore. Association for Computational Linguistics.

\bibitem[{Xiao et~al.(2021)Xiao, Hu, Liu, Tu, and Sun}]{xiao2021lawformerpretrainedlanguagemodel}
Chaojun Xiao, Xueyu Hu, Zhiyuan Liu, Cunchao Tu, and Maosong Sun. 2021.
\newblock \href {https://arxiv.org/abs/2105.03887} {Lawformer: A pre-trained language model for chinese legal long documents}.
\newblock \emph{Preprint}, arXiv:2105.03887.

\bibitem[{Xiao et~al.(2018)Xiao, Zhong, Guo, Tu, Liu, Sun, Feng, Han, Hu, Wang, and Xu}]{xiao2018cail2018largescalelegaldataset}
Chaojun Xiao, Haoxi Zhong, Zhipeng Guo, Cunchao Tu, Zhiyuan Liu, Maosong Sun, Yansong Feng, Xianpei Han, Zhen Hu, Heng Wang, and Jianfeng Xu. 2018.
\newblock \href {https://arxiv.org/abs/1807.02478} {Cail2018: A large-scale legal dataset for judgment prediction}.
\newblock \emph{Preprint}, arXiv:1807.02478.

\bibitem[{Zeng et~al.(2022)Zeng, Attarian, Ichter, Choromanski, Wong, Welker, Tombari, Purohit, Ryoo, Sindhwani, Lee, Vanhoucke, and Florence}]{zeng2022socraticmodelscomposingzeroshot}
Andy Zeng, Maria Attarian, Brian Ichter, Krzysztof Choromanski, Adrian Wong, Stefan Welker, Federico Tombari, Aveek Purohit, Michael Ryoo, Vikas Sindhwani, Johnny Lee, Vincent Vanhoucke, and Pete Florence. 2022.
\newblock \href {https://arxiv.org/abs/2204.00598} {Socratic models: Composing zero-shot multimodal reasoning with language}.
\newblock \emph{Preprint}, arXiv:2204.00598.

\bibitem[{Zhong et~al.(2018{\natexlab{a}})Zhong, Guo, Tu, Xiao, Liu, and Sun}]{zhong-etal-2018-legal}
Haoxi Zhong, Zhipeng Guo, Cunchao Tu, Chaojun Xiao, Zhiyuan Liu, and Maosong Sun. 2018{\natexlab{a}}.
\newblock \href {https://doi.org/10.18653/v1/D18-1390} {Legal judgment prediction via topological learning}.
\newblock In \emph{Proceedings of the 2018 Conference on Empirical Methods in Natural Language Processing}, pages 3540--3549, Brussels, Belgium. Association for Computational Linguistics.

\bibitem[{Zhong et~al.(2018{\natexlab{b}})Zhong, Xiao, Guo, Tu, Liu, Sun, Feng, Han, Hu, Wang, and Xu}]{zhong2018overviewcail2018legaljudgment}
Haoxi Zhong, Chaojun Xiao, Zhipeng Guo, Cunchao Tu, Zhiyuan Liu, Maosong Sun, Yansong Feng, Xianpei Han, Zhen Hu, Heng Wang, and Jianfeng Xu. 2018{\natexlab{b}}.
\newblock \href {https://arxiv.org/abs/1810.05851} {Overview of cail2018: Legal judgment prediction competition}.
\newblock \emph{Preprint}, arXiv:1810.05851.

\bibitem[{Zhong et~al.(2020)Zhong, Guo, Tu, Xiao, Liu, and Sun}]{zhong2020how}
Hongyu Zhong, Zhipeng Guo, Cunchao Tu, Chaojun Xiao, Zhiyuan Liu, and Maosong Sun. 2020.
\newblock \href {https://aclanthology.org/2020.acl-main.466} {How does nlp benefit legal system: A summary of legal artificial intelligence}.
\newblock In \emph{Proceedings of the 58th Annual Meeting of the Association for Computational Linguistics}, pages 5218--5230.

\end{thebibliography}

\appendix
\onecolumn
\section{Appendix}
\label{sec:appendix}
\begin{figure}[htbp]
    \centering
    \includegraphics[width=0.6\textwidth]{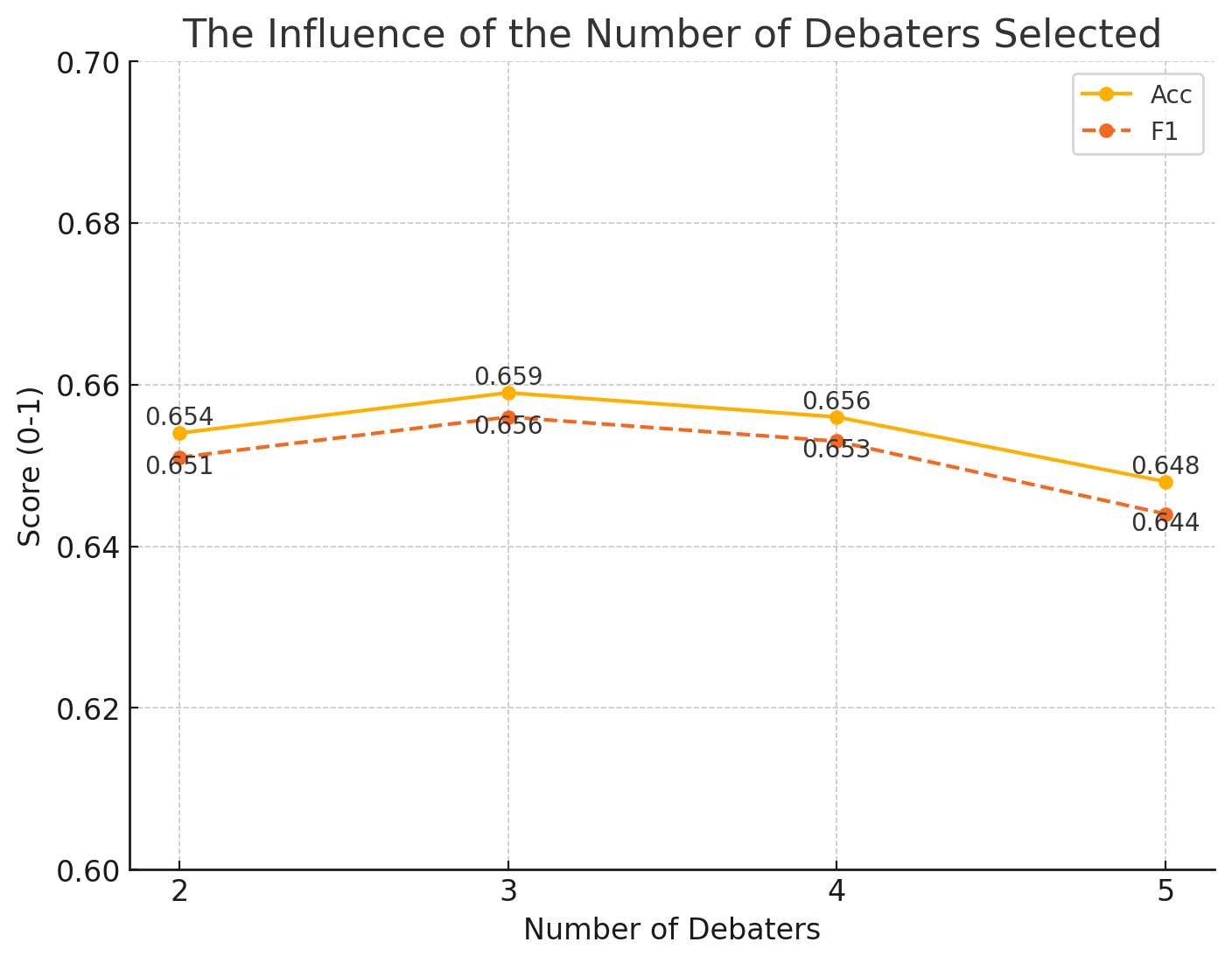}
    \caption{Influence of the number of debaters selected.}
    \label{fig:numberofdebaters} 
\end{figure}

\begin{figure}[htbp]
    \centering
    \includegraphics[width=0.6\textwidth]{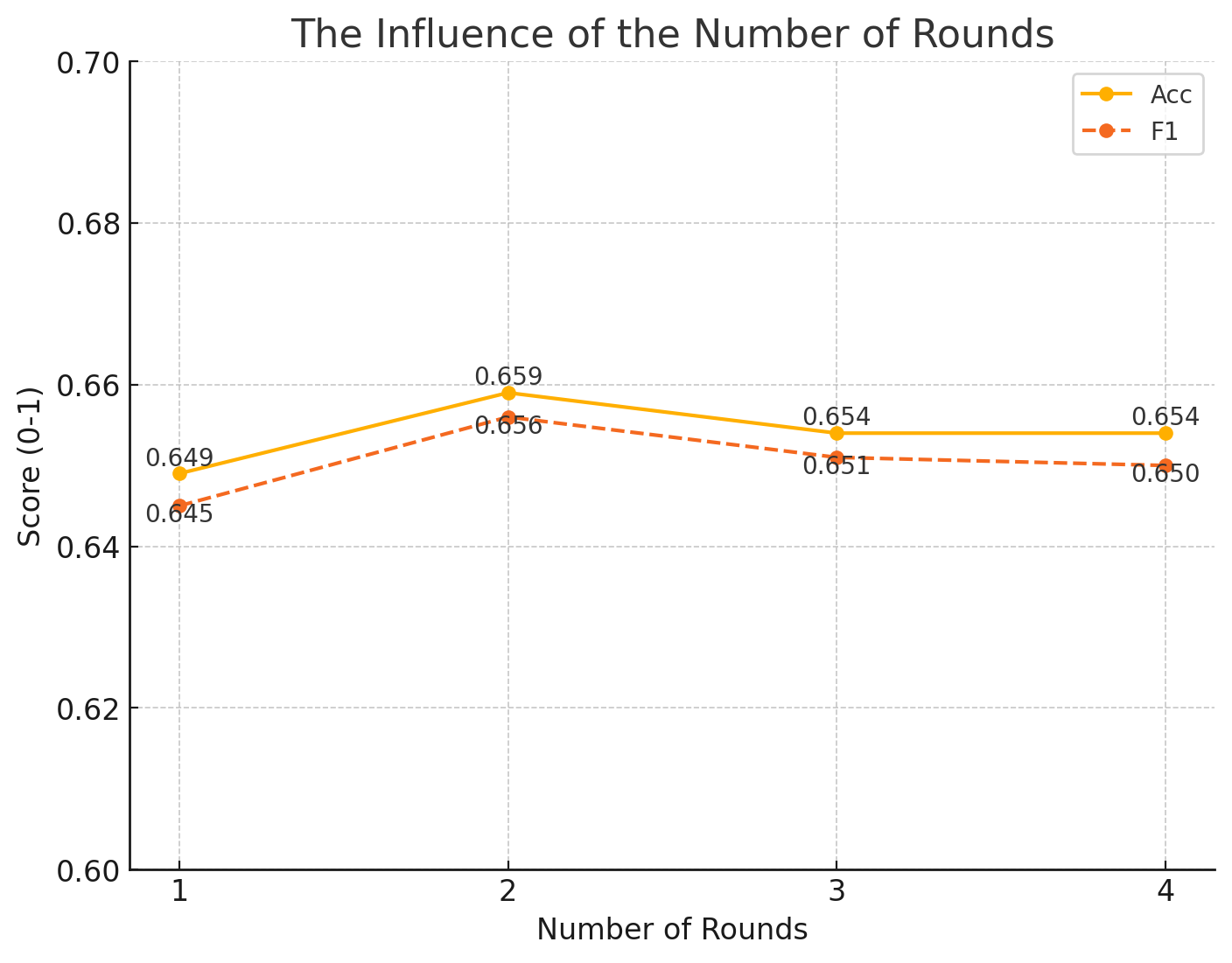}
    \caption{Influence of the number of rounds selected.}
    \label{fig:numberofrounds} 
\end{figure}

1. The choices about different numbers of rounds and debaters on the debate-feedback model (without assistant model).
\newline
\newline
As illustrated in Figures[\ref{fig:numberofdebaters}] and Figures[\ref{fig:numberofrounds}], while the number of debaters and debate rounds may vary depending on the specific task, generally, using 2-4 debaters and conducting 2-3 rounds often yields favorable results. This configuration can serve as a useful reference for readers, helping to avoid unnecessary computational overhead.

2. A sample of Debate-Feedback Structure with one round and three debaters in binary classification task, table[\ref{tab: sample of DF model}].

\small
\begin{longtable}{lll}
\small
\scalebox{0.8}{
\begin{tabular}{lll}

\hline
\multicolumn{3}{l}{1. Initial Step} \\ \hline
\multicolumn{3}{|l|}{\begin{tabular}[c]{@{}l@{}}\# input: \\ Here is a legal case, your task is to predict the result of the judge, the labels can be one of the following: \\ {[}\textbackslash{}'Plaintiff wins\textbackslash{}',\textbackslash{}'Defendant wins\textbackslash{}'{]}.\\ ONLY output the one of the labels without anything else.\\ Here is the case background:\\ The case, "The Adams Express Company v. Richard T. Haynes," took place in 1865, involving a transaction where Richard T. Haynes entrusted a \\ trunk to the Adams Express Company for transport from Nashville, Tennessee, to Chicago, Illinois. Haynes claimed that the contents of the \\ trunk were lost due to the negligence and fraudulent actions of the express company. The express company provided a receipt outlining the \\ terms of service, which limited their liability ... "\end{tabular}} \\ \hline
\rowcolor[HTML]{FFFFFF} 
\multicolumn{3}{|l|}{\cellcolor[HTML]{FFFFFF}\begin{tabular}[c]{@{}l@{}}\# output:\\ Plaintiff wins\end{tabular}} \\ \hline
\rowcolor[HTML]{FFFFFF} 
\multicolumn{3}{|l|}{\cellcolor[HTML]{FFFFFF}2. Debate Step} \\ \hline
\multicolumn{1}{|l|}{\cellcolor[HTML]{FFFFFF}Debater 0} &
  \multicolumn{1}{l|}{Debater 1} &
  \multicolumn{1}{l|}{Debater 2} \\ \hline
\rowcolor[HTML]{FFFFFF} 
\multicolumn{1}{|l|}{\cellcolor[HTML]{FFFFFF}\begin{tabular}[c]{@{}l@{}}\# input: \\ Here is a legal case, your task is to \\ predict the result of the judge and \\ Debate withother experts, the labels \\ can be one of the following: \\ {[}\textbackslash{}'Plaintiff wins\textbackslash{}',\textbackslash{}'Defendant wins\textbackslash{}'{]}. \\ Please state your opinion and refer to \\ factual examples whenever possible.\\ Your answer no more than 200 words.\\ Here is the case background:\\ ...\\ \\ \# output:\\ I predict that the judge will rule in \\ favor of the plaintiff, Richard T. Haynes. \\ Although the express company presented a \\ receipt with terms limiting its liability, \\ the crux of the case centers on the \\ allegations of gross negligence and possible \\ fraudulent actions ...\end{tabular}} &
  \multicolumn{1}{l|}{\cellcolor[HTML]{FFFFFF}\begin{tabular}[c]{@{}l@{}}\# input: \\ Here is a legal case, your task is to \\ predict the result of the judge and \\ Debate withother experts, the labels \\ can be one of the following: \\ {[}\textbackslash{}'Plaintiff wins\textbackslash{}',\textbackslash{}'Defendant wins\textbackslash{}'{]}. \\ Please state your opinion and refer to\\ factual examples whenever possible.\\ Your answer no more than 200 words.\\ Here is the case background:\\ ...\\ \\ \# output\\ In the case of "The Adams Express Company\\  v. Richard T. Haynes," it is likely that \\ the plaintiff, Haynes, would prevail due to\\ the allegations of gross negligence and fraud.\\ Despite the express company\textbackslash{}'s attempt to \\ enforce the terms of limited liability, these \\ clauses typically ...\end{tabular}} &
  \multicolumn{1}{l|}{\cellcolor[HTML]{FFFFFF}\begin{tabular}[c]{@{}l@{}}\# input:\\ Here is a legal case, your task is to \\ predict the result of the judge and \\ Debate withother experts, the labels \\ can be one of the following: \\ {[}\textbackslash{}'Plaintiff wins\textbackslash{}',\textbackslash{}'Defendant wins\textbackslash{}'{]}. \\ Please state your opinion and refer to \\ factual examples whenever possible.\\ Your answer no more than 200 words.\\ Here is the case background:\\ ...\\ \\ \# output\\ In the case of "The Adams Express Company \\ v. Richard T. Haynes," I predict that the \\ judge will rule in favor of the plaintiff, \\ Richard T. Haynes. The key issue revolves \\ around whether the express company\textbackslash{}'s \\ actions constituted gross negligence or \\ fraud, which would negate their ...\end{tabular}} \\ \hline
\rowcolor[HTML]{FFFFFF} 
\multicolumn{1}{|l|}{\cellcolor[HTML]{FFFFFF}\begin{tabular}[c]{@{}l@{}}\# input \\ Now I will give you the opinion of another \\ expert, if their opinions differ from yours,\\ you can support their views or offer a \\ counter-argument.\\ This round of conversation will be provided \\ to the judge for reference, so remember to \\ express your stance clearly.\\ Your answer no more than 200 words.\\ Here are their opinions:\\ ...\\ \\ \# output\\ I largely agree with the perspective \\ of the first expert regarding the implications \\ of gross negligence and fraud in "The Adams \\ Express Company v. Richard T. Haynes." The\\  pivotal issue is whether Haynes can \\ convincingly demonstrate\\ ...\end{tabular}} &
  \multicolumn{1}{l|}{\cellcolor[HTML]{FFFFFF}\begin{tabular}[c]{@{}l@{}}\# input\\ Now I will give you the opinion of another \\ expert, if their opinions differ from yours, \\ you can support their views or offer a \\ counter-argument.\\ This round of conversation will be provided \\ to the judge for reference, so remember to \\ express your stance clearly.\\ Your answer no more than 200 words.\\ Here are their opinions:\\ ...\\ \\ \# output\\ I strongly agree with the assessment of the \\ first expert regarding the likelihood of the \\ plaintiff, Richard T. Haynes, prevailing in \\ "The Adams Express Company v. Richard T. Haynes." \\ The focus on the express company\textbackslash{}'s \\ ...\end{tabular}} &
  \multicolumn{1}{l|}{\cellcolor[HTML]{FFFFFF}\begin{tabular}[c]{@{}l@{}}\# input\\ Now I will give you the opinion of another \\ expert, if their opinions differ from yours, \\ you can support their views or offer a \\ counter-argument.\\ This round of conversation will be provided \\ to the judge for reference, so remember to\\ express your stance clearly.\\ Your answer no more than 200 words.\\ Here are their opinions:\\ ...\\ \\ \# output\\ I find that I am in strong agreement with the \\ opinion of the second expert regarding the case \\ of "The Adams Express Company v. Richard T. \\ Haynes." The emphasis on the express company\textbackslash{}'s \\ potential failure to demonstrate \\ ...\end{tabular}} \\ \hline
\rowcolor[HTML]{FFFFFF} 
\multicolumn{3}{|l|}{\cellcolor[HTML]{FFFFFF}3. Verification Step} \\ \hline
\rowcolor[HTML]{FFFFFF} 
\multicolumn{1}{|l|}{\cellcolor[HTML]{FFFFFF}\begin{tabular}[c]{@{}l@{}}\{input: the summary of Debater 0's statement\\  to the assistant model\}\\ \{output: reliability: 0.75\}\end{tabular}} &
  \multicolumn{1}{l|}{\cellcolor[HTML]{FFFFFF}\begin{tabular}[c]{@{}l@{}}\{input: the summary of Debater 1's statement \\ to the assistant model\}\\ \{output: reliability: 0.8\}\end{tabular}} &
  \multicolumn{1}{l|}{\cellcolor[HTML]{FFFFFF}\begin{tabular}[c]{@{}l@{}}\{input: the summary of Debater 2's statement \\ to the assistant model\}\\ \{output: reliability: 0.45\}\end{tabular}} \\ \hline
\rowcolor[HTML]{FFFFFF} 
\multicolumn{3}{|l|}{\cellcolor[HTML]{FFFFFF}4. Output step} \\ \hline
\rowcolor[HTML]{FFFFFF} 
\multicolumn{3}{|l|}{\cellcolor[HTML]{FFFFFF}\begin{tabular}[c]{@{}l@{}}\# input:\\ \{Follow the conversation from Step 1\}\\ Some experts discussed the case in this round and here is the summary, please provide your final opinion.\\ Same as above, ONLY output the one of the labels without anything else.\\ Here is the debate summary:\\ {[}\textbackslash{}'In the case of ...{]}\\ \{insert the reliability from Step 3\}\\ \\ \# output \\ Plaintiff wins\end{tabular}} \\ \hline
\caption{An Example of Debate-Feedback Structure}
\label{tab: sample of DF model}
\end{tabular}
}

\end{longtable}{}
\normalsize
3. Performance of the smoothing mechanism.
\newline
\newline

\begin{table}[h]
    \centering
    \scalebox{0.8}{
    \begin{tabular}{|l|c|c|c|}
        \hline
        \textbf{Debate-Feedback Mechanism} & \textbf{Prediction Correction} & \textbf{Prediction Degradation} & \textbf{Accuracy Rate} \\
        \hline
        Without Smoothing & 102 & 115 & 62.8\% \\
        \hline
        With Smoothing & 93 & 11 & 65.7\% \\
        \hline
    \end{tabular}
    }
    \caption{Performance of smoothing mechanism.}
    \label{tab:debate_feedback}
\end{table}

In our initial experiments, we unexpectedly discovered that a simple smoothing operation was particularly useful in improving prediction accuracy. Specifically, we tested the Prediction Correction Rate and Prediction Degradation Rate with and without smoothing on a binary CaseLaw dataset containing 3000 samples, as shown in table[\ref{tab:debate_feedback}].

\begin{itemize}
    \item \textbf{Prediction Correction}: When the initial prediction of the model is wrong, and it is corrected by the debate-feedback framework.
    \item \textbf{Prediction Degradation}: When the initial prediction of the model is correct, but becomes incorrect due to the framework.
\end{itemize}

We found that the Prediction Degradation Rate was particularly high without smoothing, while the Prediction Correction Rate was about the same. This means the smoothing mechanism helps models avoid relying too heavily on the influence of a certain debater.

\end{document}